# Temperature and Electric Field Induced Metal-Insulator Transition in Atomic Layer Deposited Vanadium Dioxide Thin Films

**Marko J. Tadjer[1,a], Virginia D. Wheeler[1], Brian Downey[1], Zachary R. Robinson[2], David J. Meyer[1], Charles R. Eddy, Jr.[1], Fritz J. Kub[1]**

1 – United States Naval Research Laboratory, Washington DC, USA
2 – Department of Physics, The College at Brockport, SUNY, NY, USA
a – marko.tadjer@nrl.navy.mil

**Abstract**

Amorphous vanadium oxide ($VO_2$) films deposited by atomic layer deposition (ALD) were crystallized with an ex situ anneal at 660-670 °C for 1-2 hours under a low oxygen pressure ($10^{-4}$ to $10^{-5}$ Torr). Under these conditions the crystalline $VO_2$ phase was maintained, while formation of the $V_2O_5$ phase was suppressed. Electrical transition from the insulator to the metallic phase was observed in the 37-60 °C range, with a $R_{ON}/R_{OFF}$ ratio of up to about 750 and $\Delta T_C \cong$ 7-10 °C. Electric field applied across two-terminal device structures induced a reversible phase change, with a room temperature transition field of about 25 kV/cm in the $VO_2$ sample processed with the 2 hr long anneal. Both the width and slope of the field induced MIT hysteresis were dependent upon the $VO_2$ crystalline quality.

**Keywords**

Vanadium dioxide, $VO_2$, metal insulator transition, phase change, thermochromic, conductivity, resistivity, strongly correlated electrons.

**Introduction**

Vanadium oxide has several crystalline phases, many of which undergo a temperature dependent structural phase change [1-4]. In particular, the VO$_2$ phase exhibits the metal-insulator transition (MIT) ideally at a critical transition temperature (T$_C$) of ~68 °C [5-8]. This structural shift from the monoclinic phase to the tetragonal rutile phase results in a significant change in electrical conductivity, optical transmittance and reflectance, as well as thermal emittance [9, 10]. Due to these temperature dependent properties, VO$_2$ films have been of significant interest for applications such as microbolometers, passive spacecraft thermal management, RF switches, Mott memristors, field-effect transistors, and other novel concepts for switching devices [11-16]. For such applications, atomic layer deposition (ALD) is an important integration method offering superior wafer-scale thickness control and enabling ultrathin films and conformal coating of complex surfaces and architectures. A wafer-scale conformal ALD VO$_2$ film with the capability to transition to metallic state at room temperature under applied electric field has the potential to operate at a commercial scale without additional heating systems in order to maintain the device near the transition temperature. Such properties could also be desirable in future space applications where temperature control system integration is undesirable or prohibitively expensive. In addition, field effect transistors based on VO$_2$ have already greatly benefitted from high-k dielectrics such as HfO$_2$ [16]. In such novel devices, even better field-effect response could potentially be obtained if the entire device structure was to be grown by ALD and tuned by an in situ anneal.

Achieving MIT for very thin (up to 50 nm) $VO_2$ films has been challenging even after annealing [17]. The interplay of MIT-induced delocalized (correlated) electrons in the metallic state leads to a high sensitivity of the conductivity to external influences such as temperature and field, and possibly pressure [6, 7]. Both temperature and voltage dependent MIT in ALD-deposited films have been demonstrated, however, a systematic characterization of switching characteristics in such thin films as a function of $VO_2$ crystallization conditions has not been performed to date [18]. The temperature dependence of electric field induced MIT in $VO_2$ has been attributed to Joule heating by several authors [19-23]. In addition, field-induced switching has also been achieved in the absence of Joule heating using transverse field applied using either a gated field effect structure or electric force microscopy [16, 21]. In this work, we demonstrate that ALD-deposited amorphous $VO_2$, when annealed to the desired degree of crystallinity, is capable of field-induced MIT with a transitioning voltage in excess of 10 V at 30 °C, as well as at a transition at higher temperatures up to the critical temperature of transition (Tc). We demonstrate that 50 nm thick $VO_2$ films with resistivity in the monoclinic and rutile phases ($R_{OFF}$ and $R_{ON}$, respectively) could exhibit comparable properties to those of thicker (> 100 nm) films deposited by magnetron sputtering or pulsed laser deposition (PLD) [6, 7].

**Experimental Procedure**

Amorphous, 50 nm thick $VO_2$ films were deposited on double-side polished c-plane sapphire substrates by ALD in an Ultratech Savannah 200 reactor at 150 °C with a growth rate of 0.9 Å/cycle using tetrakis(ethylmethyl)amido vanadium (TEMAV, Air

Liquide) and ozone precursors. In order to achieve crystallinity, ALD $VO_2$ films grown under identical conditions were subsequently annealed in an oxygen-rich environment in a controlled $O_2$ partial pressure in the $10^{-4}$-$10^{-5}$ Torr range. Samples A and B were annealed at $P_{O2} = 10^{-5}$ Torr for 1 hr at a temperature of 660 °C and 2 hrs at 670 °C, respectively, as measured by a thermocouple attached to the bottom side of the sample stage. Sample C was annealed at $P_{O2} = 10^{-4}$ Torr for 1 hr at 660 °C (see Table I).

The differences in annealing time and temperature resulted in different structural properties as determined by powder X-ray diffraction (XRD) and atomic force microscopy (AFM). The effect of growth conditions and subsequent annealing conditions on the structural and chemical and thus electrical and optical transition properties of the $VO_2$ films will be extensively reported elsewhere [24]. In the present experiment, we designed our sample set in order to systematically explore the relationship between film structure and the resulting electrical characteristics.

| Sample | A | B | C |
|---|---|---|---|
| Annealing Temperature (°C) | 660 | 670 | 660 |
| Annealing Time (hr) | 1 | 2 | 1 |
| $O_2$ Partial Pressure (Torr) | 1e-5 | 1e-5 | 1e-4 |
| Surface roughness rms (nm) | 1.1 | 1.3 | 1.36 |
| $E_I$ (eV) | 0.63 | 0.22 | 0.91 |
| $E_T$ (eV) | 0.98 | 2.89 | - |
| $E_M$ (eV) | 0.12 | 0.36 | - |
| $T_C$ (°C) | 37 | 51 | 60 |
| Temperature Hysteresis, $\Delta T_C$ (°C) | 7 | 10 | 8 |
| $R_{OFF}/R_{ON}$ | 15 | 200 | 750 |

Table I. Summary of relevant parameters and activation energies in the insulator ($E_I$), transition ($E_T$), and metallic ($E_M$) states of samples A and B.

Devices with active area of 4-12 μm by 130 μm were patterned using an $SF_6$-based inductively-coupled plasma (ICP) process (5 mT, 200 W ICP, 50 W RIE, 40 sccm $SF_6$, 45 s). Two-terminal crystalline $VO_2$ test structures were fabricated with 20/200 nm Ti/Au contacts for characterization of electrical MIT properties. A KOH-free DUV photoresist developer (MIBK series) was used in the lithography process in order to avoid etching the $VO_2$. Electrical measurements were performed in air using an HP4145B parameter analyzer and a temperature-controlled contact probe station with an integrated low-noise MDC 490 DC heater with 1 °C accuracy.

**Results and Discussion**

AFM measurements indicated the surface roughness of the annealed $VO_2$ samples, summarized in Table I, at least a factor of 2 lower than $VO_2$ grown by other groups [9, 22]. The slightly rougher surfaces of samples B and C correlated with improved crystallinity, as shown by the XRD data in Fig. 1. All three samples exhibited a polycrystalline $VO_2$ phase with peaks indicative of the (110) and (020) orientations [25]. The intensity of the (110) reflection at 27.8° increased with increasing annealing temperature and time as the oxygen partial pressure was kept at $10^{-5}$ Torr (samples A and B). On the other hand, the (020) reflection at 39.8° was highest when the annealing temperature and time were 1 hr and 660 °C, respectively, but the $O_2$ partial pressure was higher (sample C, $10^{-4}$ Torr). The observed improvement in $VO_2$ crystallinity by increasing the annealing time, temperature, or $O_2$ pressure were correlated to the electrical characteristics of the two-terminal test structures.

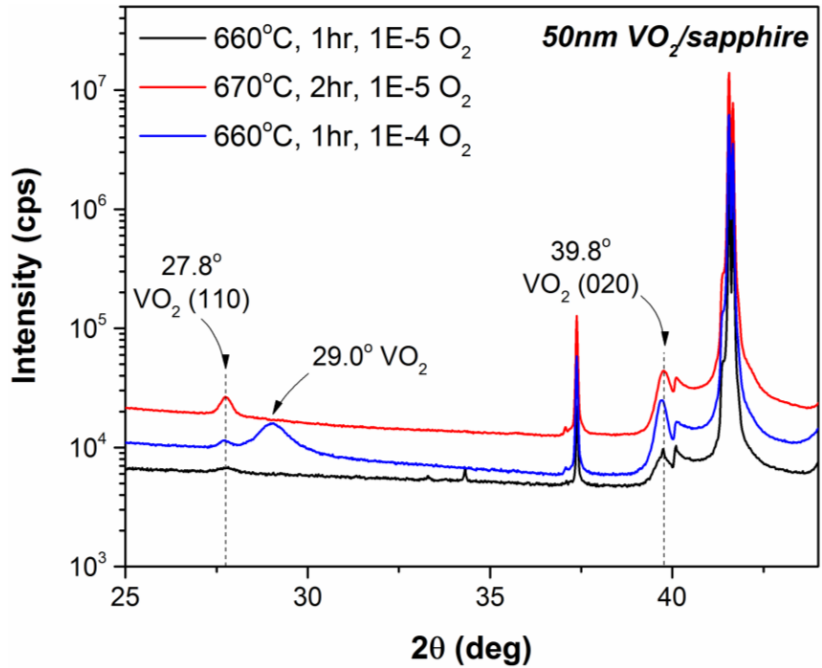

Fig. 1. Powder X-ray diffraction spectra for 50 nm thick ALD deposited $VO_2$ films on c-$Al_2O_3$ and annealed ex situ at $10^{-5}$ Torr pressure in $O_2$ at 660 °C for 1 hr (Sample A, black) and 670 °C for 2 hrs (Sample B, red), respectively, as well as at $10^{-4}$ Torr pressure at 660 °C for 1 hr (Sample C, blue).

Sheet resistance was measured via the transfer length method as a function of temperature and is presented in Fig. 2. Sample C exhibited a $T_C$ closest to the theoretical value of 68 °C, sharpest MIT, highest on-state and off-state resistances, as well as highest $R_{ON}/R_{OFF}$ ratio (see Table I), all of which are indications that the higher oxygen pressure used during the anneal of sample C resulted in a higher crystallinity $VO_2$ film than obtained at a longer annealing time for sample B. This result was in good agreement with the XRD data in Fig. 1.

From Fig. 2, $T_C$ was estimated to be 40-50 °C for sample A 55-65 °C for sample B, and 60-70 °C for sample C, due to the hysteresis induced by the temperature cycle. Sample B exhibited a slightly larger hysteresis than samples A and C ($\Delta T_C$, Table I). Unlike sample A, the slope of the transition for samples B and C in Fig. 2 was independent of the direction of temperature change, possibly indicating that the temperature-dependent I-V measurements additionally crystallized the VO$_2$ film in sample A.

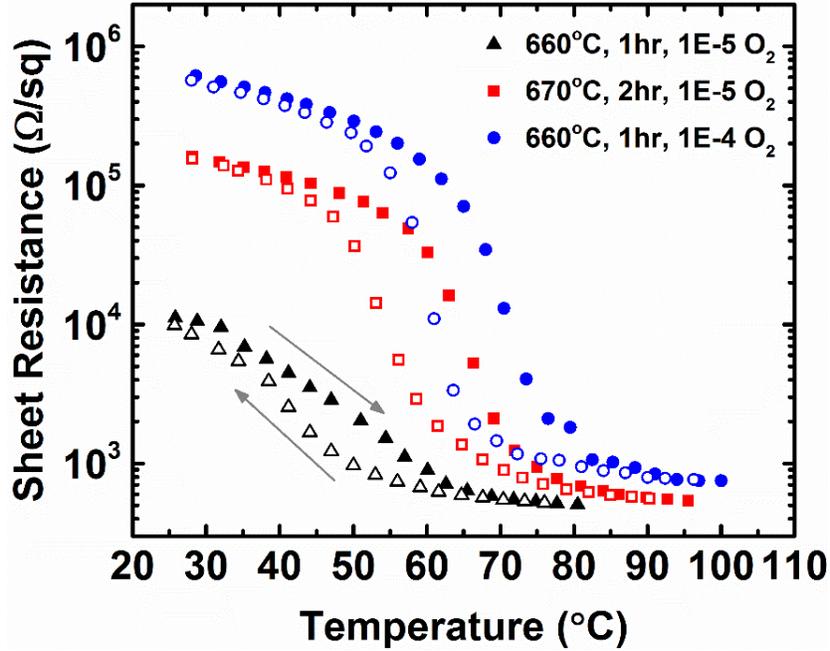

Fig. 2. Sheet resistance of samples A (▲), B (■), and C (●) measured as a function of temperature, showing reversible MIT in all three samples. Sample C exhibited highest transition temperature ($T_C$), off-state resistance ($R_{OFF}$), on-state resistance ($R_{ON}$), and $R_{OFF}/R_{ON}$ ratio.

Current-voltage characteristics from the three samples were measured from -20 V to +20 V and back to -20 V at 50 mV steps. All measurement cycles were performed from 30 °C to 80 °C range at 1 °C increments. Figure 3 presents the I-V-T characteristics from test structures with 4 μm contact separation, and includes the current response when the bias was increased from the insulating to the metallic state (i.e., from 0 V towards +20 V and from 0 V towards -20 V). The current measured from samples A and B during a complete voltage sweep at 30 °C is given in Fig. 4a. Similar reversible transitions from the metallic state back to the insulating state, confirming the non-catastrophic nature of all electrical measurements, were obtained from these samples up to 80 °C.

A sharp increase in current, was measurable across the entire temperature range in samples A and B (Fig. 3a-b), whereas only a small change in current was measured in sample C under identical measurement conditions (Fig. 3c). While sample C was shown to be of highest crystalline quality and exhibited the best thermally induced electrical characteristics (e.g., highest $R_{OFF}/R_{ON}$ ratio), it was not optimal for field-induced phase transition. MIT at a bias higher than 20 V was measurable in sample C, however, at higher bias the reliability of the contacts degraded and the measured current did not controllably transition as it did with samples A and B. We thus assumed that contact reliability for $VO_2$ devices is a challenging task since any contact resistance reduction through annealing must be performed under conditions that are benign for the properties of the $VO_2$ films.

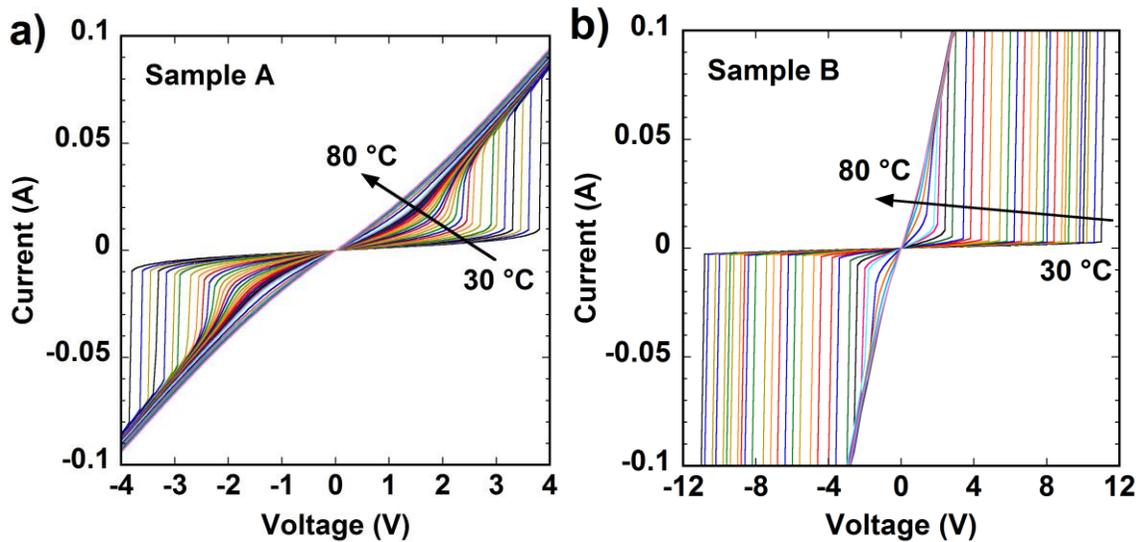

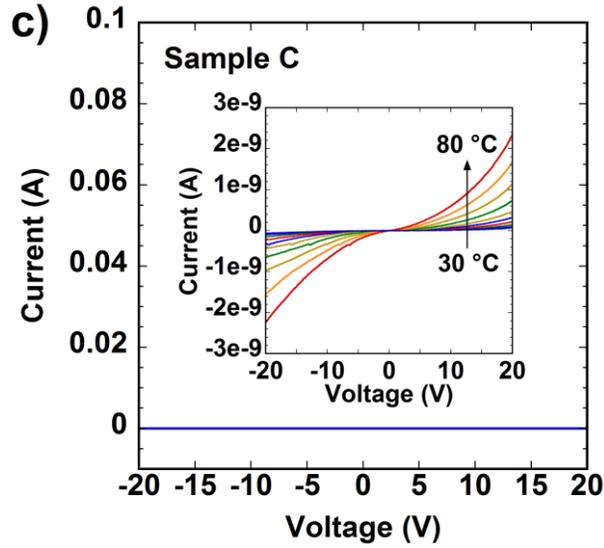

Fig. 3. Temperature-dependent I-V measurements showing a sharp reversible MIT for (a) sample A and (b) sample B, and no MIT for (c) sample C within the 30-80°C and ±20 V measurement ranges.

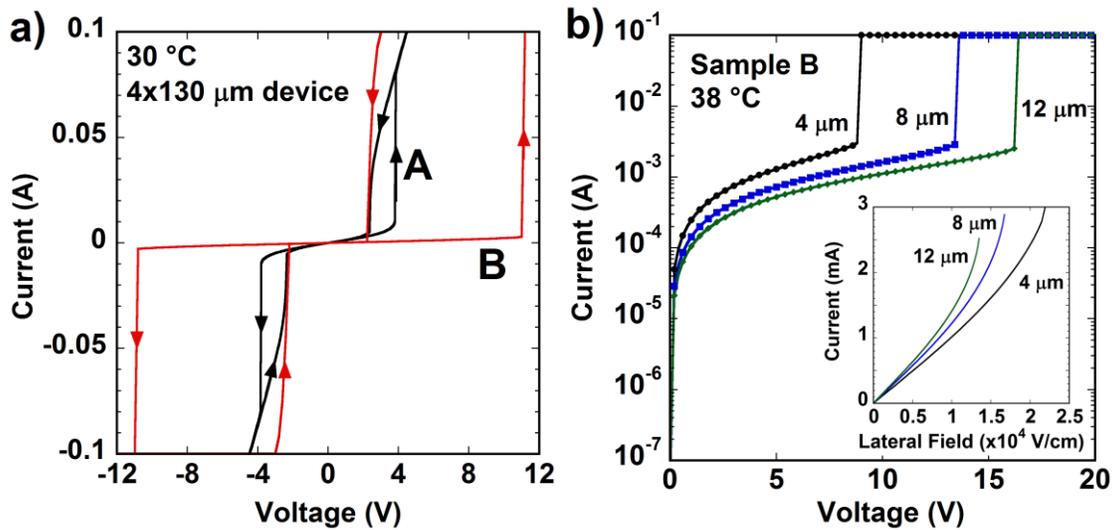

Fig. 4. (a) Electric field induced hysteresis comparison at 30 °C for samples A (black) and B (red). (b) MIT field transition as a function of contact spacing at 38 °C for sample B. Inset: data from (b) normalized to lateral field across the film on the x-axis.

As presented in Fig. 3, the higher crystalline quality of sample B correlated well with the higher transition field, compared to sample A. Similarly, the sweep direction dependent hysteresis profiles in Fig. 4a revealed more than 5 times wider transition voltage in sample B. While the temperature dependent hysteresis data was not provided for brevity, the hysteresis area for both samples decreased as the temperature was increased, similarly to prior reports [22]. The power needed for MIT in sample B decreased linearly with temperature, confirming Joule heating was the predominant switching mechanism for that sample [22]. However, this was not the case for sample A, where the $VO_2$ film was partially crystallized and a soft (non-abrupt) transition profile was measurable even near room temperature in the 1.8-2.5 V range. Nonlinear current was also measurable in sample B in this voltage range but in a temperature range of only about 5 °C. We have attributed the behavior in this "soft-MIT" region to the fact that the thermal conductivity of $VO_2$ is also phase dependent and could change by as much as 60 percent over the course of the MIT, ultimately leading to a non-abrupt field switching profile as the critical temperature is approached during measurement [26]. We also note that the softer field-induced transition profile of sample A correlated with the smaller slope of the temperature induced MIT for this sample, as shown in Fig. 2.

The hysteresis area increased with device spacing, as indicated by the higher transition fields in Fig. 4b. The measurement temperature of 38 °C was the lowest temperature where field-assisted MIT was observed for contact separations up to 12 μm, up to the 100 V limit of the instrument. The inset shows that tripling the contact spacing did not increase the transition field by a factor of three. Thus, the highest field of MIT was measured on the shortest $VO_2$ devices, an effect likely due to second-order effects such as heat spreading through the contacts [18]. The highest MIT transition field for sample B was ~25 kV/cm and was measured at 30 °C with a 4 μm contact separation. For comparison, previous reports of field-induced MIT in 100 nm thick crystalline ALD $VO_2$ showed a similar transition field (determined by the energy gap of crystalline $VO_2$), but measured over a much smaller active area device (0.2 MV/cm over a 0.1x10 μm device) [18].

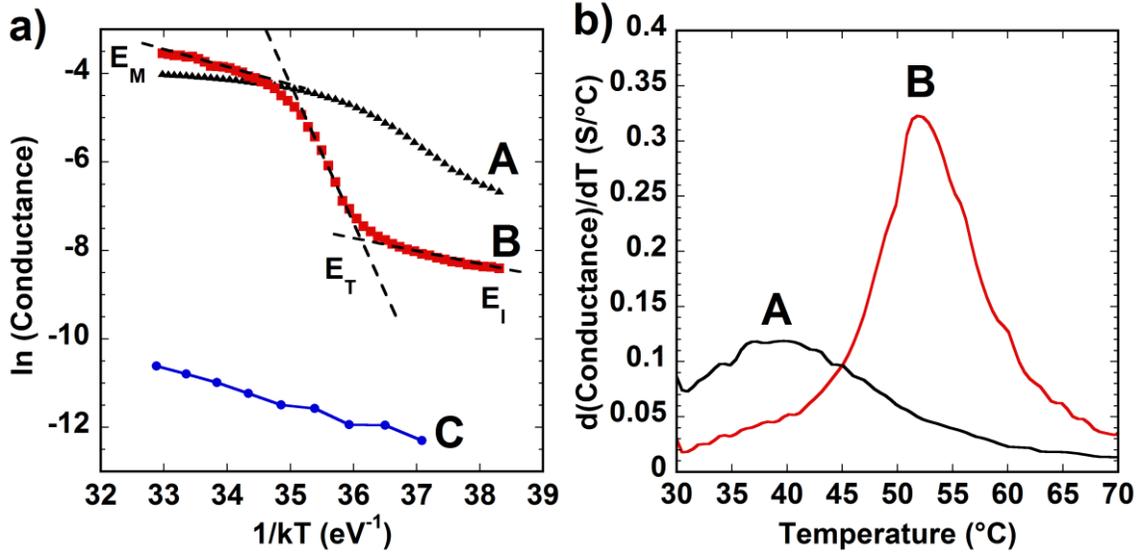

Fig. 5. (e) Arrhenius plot of the temperature-dependent conductance and (f) derivative of the conductance with respect to temperature for samples A (black) and B (red).

The conductance of each sample was extracted from the linear portion of the I-V data and plotted on an Arrhenius graph in Fig. 5a. Activation energy was thus possible to extract for the applicable $VO_2$ state for each sample (insulating, transitioning, metallic) and was provided in Table I. The peak of the first derivative of the conductance (Fig. 5b) was used to more accurately determine the $T_C$ in samples A and B (37 and 51 °C, respectively). The 0.91 eV insulating state activation energy of sample C originated from its non-transitioning behavior across the entire measured temperature range. The higher transition-state activation energy ($E_T$) for sample B correlated with a higher $T_C$, and the higher metallic-state activation energy ($E_M$) correlated with a higher field (higher power, thus higher Joule heating) required for MIT.

The different activation energies corresponding to each structural phase of $VO_2$ indicate that both temperature and field dependent conductivity would be highly dependent on the particular crystallinity of the annealed $VO_2$. We point out that electrical data alone cannot definitively point to Joule heating over electron delocalization due to structural transition as the sole driver for field-induced MIT, especially in the soft MIT region noticeably measureable in sample A. Such a separation of mechanisms would be further complicated by practical challenges in localized temperature measurements of $VO_2$ devices under bias such as heat spreading, strain, and nonuniform thermal conductivity.

**Conclusions**

In 50 nm thick $VO_2$ films with an active area of up to 1560 $\mu m^2$, MIT at room-temperature with transition field in excess of 20 kV/cm (sample B), as well as controllable MIT profiles based on careful post-growth annealing conditions, were demonstrated. The higher temperature and field needed for transition in sample B suggested that the (110) $VO_2$ orientation was preferable for electric field induced switching. On the other hand, the (020) crystalline orientation resulted in the best resistivity-temperature profile (sample C), however, electric field induced MIT was suppressed. Thus, we demonstrate that the structural properties of $VO_2$, which determine the resistivity profile and are already sensitive to multiple factors such as doping and strain, must be carefully tuned in order to obtain the desired electrical response [27, 28].

**Acknowledgements**

The authors are sincerely grateful to the following NRL staff: Mr. Neil Green for sample processing, Dr. Konrad Bussman for annealing equipment support, and Dr. Karl Hobart for insightful discussions.